\documentclass{article}
\usepackage[a4paper, bottom=1.2in, top=1.2in, left=1.2in, right=1.2in, bindingoffset=1cm]{geometry}

\usepackage{amsfonts}
\usepackage{amssymb}
\usepackage{amsmath}
\usepackage{amsthm}
\usepackage{mathtools}
\usepackage{accents}
\usepackage{graphicx}
\usepackage{caption}
\usepackage{subcaption}
\usepackage{tikz}
\usepackage[hidelinks]{hyperref}
\usepackage{pgfplots}
\usepackage{filecontents}

\usepackage{xcolor}
\hypersetup{
	colorlinks,
	linkcolor={white!15!black},
	citecolor={white!15!black},
	urlcolor={white!20!black}
}

\usepackage[ruled,vlined,procnumbered,noend]{algorithm2e}

\usetikzlibrary{decorations.pathreplacing,shapes,calc,angles,quotes}
\usepgfplotslibrary{patchplots}

\newtheorem{theorem}{Theorem}

\newtheorem{fact}{Fact}

\newtheorem{definition}{Definition}[theorem]

\newcommand{\NN}{\mathbb{N}}

\title{ Preliminary Report: On Information Hiding in Multi-Hop Radio Networks}
\author{Marek Klonowski, Mateusz Marciniak}
\date{}

\begin{document}
	\maketitle
	
	\begin{abstract}
	In this paper, we consider the problem of an adversary aiming to learn information about the network topology or the executed algorithm from some signals obtained during the algorithm's execution. The problem is defined in a very general form. However, it is mainly motivated by multi-hop ad hoc radio networks. In contrast to previous work concentrated on single-hop radio networks, this model is critically more complex due to the number of possible settings that need to be taken into account when considering different combinations of topologies and communication models. Moreover, the definition of the adversary is also ambiguous, and the adequate approach needs to depend on the adversary's aims and capabilities. This preliminary report presents a general theoretical background and some basic algorithms. We also propose some general taxonomy as a framework for future research. 
	
\end{abstract}
	\section{Introduction}\label{intro}

In this paper, we consider the problem of a distributed algorithm execution from the perspective of hiding some (meta)-information from the curious observer (an adversary) having access to some data sources related to the execution (feedback from the algorithm execution). We present the idea in possibly high-level/abstract form, covering a wide range of distributed systems. Nevertheless, to make the presentation more transparent and well-motivated, we focus on a  multi-hop, synchronous ad hoc radio network, where an adversary can observe some transmissions and possibly partially know the network's topology. The motivation behind the information hiding for this type of system is straightforward: learning the details of the protocol execution may reveal some information about inputs for the executed distributed algorithm (e.g., the number of packets processed by individual stations), the type of algorithm that is executed or properties of the underlying network (e.g., number of stations in the network). Revealing such information can be highly undesirable, e.g., a group of robots collectively exploring some terrain shall not reveal too much about the algorithm they execute, e.g., in military applications. 

We assume that the network is operating in a multi-hop model. There may be no direct communication between some pairs of devices and communication between some pairs must be conducted by some kind of relay. Such communication might also require some protection against eavesdropping - potential adversaries may infer some pieces of information about the topology of the network, having some details of protocol execution. For example, the total number of transmissions or the length of the execution of a given algorithm may be strongly correlated with the network's diameter. Moreover, in some cases, the details of the algorithm's execution allow the adversary to recover the exact topology. In the broad spectrum of considered adversarial-observer cases, we also consider the scenario wherein the adversary aims to learn the details of the executed algorithm (e.g., local inputs, type of algorithm) while the topology is known. 

To some extent, this paper can be seen as a continuation of \cite{DBLP:conf/europar/BojkoKMS23}, wherein a similar problem has been considered for a single hop radio network with the beeping model. However, the problem analyzed in that paper has been restricted only to hiding the number of stations participating in the distributed algorithm. 

In contrast to previous work concentrated on single-hop radio networks, in multi-hop settings, we need to consider various network models and different capabilities of the adversary aiming at learning the details of the execution and the network itself. Moreover, we point out that the adversary's capabilities and possible countermeasures to hide some information strongly depend on the communication model. That is, replacing the plain beeping model 
in~\cite{DBLP:conf/europar/BojkoKMS23} with another communication channel (e.g. classic noCD) may result in dramatically different analysis and results even in the single-hop case.  

This paper presents only preliminary research pointing out how complicated and versatile are various cases of hiding information in multi-hop networks. Apart from the formal model, we present a taxonomy of different models from the perspective of information hiding and a few basic protocols for just a few models.

\paragraph{Organization of this paper}
In the Section~\ref{model}, we introduce some details of the formal model. Section~\ref{taxonomy} is devoted to the taxonomy of possible network settings and adversarial models. 
In Section~\ref{algorithms}, we present some chosen algorithms. We present the most important related work in \autoref{related}.

    \section{Model}\label{model}

Describing a formal model for our problem is complex since we want to consider all essential details. First,  we need to describe the network, including settings governing synchronization and the capabilities of nodes. Then, we need to describe the communication model - how the information is transmitted using the communication channel. Finally, we need to specify the security model, in particular, the capabilities of the adversary. In particular, we need to specify the \textit{feedback function}, i.e., what the adversary learns from the protocol's execution. Let us stress that the described systems can consider different feedback functions motivated by different real-life scenarios depending on the distributed system and the acting of the adversary.

\subsection{Network model}
The network is represented by undirected, connected graph $G = (V, E)$, where $V$ will be a set of stations and $|V| = n$. Stations are connected by edges $\{v, u\} \in E$, where $u, v \in V$. When there is an edge between two stations, they can bidirectionally communicate with each other and receive/send some information through that links. Each station's $v \in V$ set of neighbours is denoted as $N(v)=\{u : \{u, v\} \in E\}$ and $N^+(v) = N(v)\cup\{v\}$. Let $D$ be the diameter of the network. 

\subsection{Communication channel}\label{qqRef}

Communication between stations will be synchronized by a global clock accessible for all stations and split into \textit{slots}. 
In each slot, a station can transmit or listen. The transmission emitted by a station $v$ reaches all neighbors of $v$, i.e., $N(v)$. We consider the following communication channels:

\begin{description} 
	\item[Beeping model] (e.g. \cite{deploying_wireless_networks_with_beeps, near_optimal_leader_election_in_multi_hop_radio_networks, optimal_multi_broadcast_with_beeps_using_group_testing, CZUMAJ20192})
	- the signal is received by a station $v$ if and only if at least one station from the set $N(v)$ is transmitting and $v$ is in the listening mode. It is the simplest model, where each station $v$ can recognize only two communication states: \textit{Beep} if any station from $N^+(v)$ transmitted a message in this round and \textit{Silence} when no station transmitted.
	
	\item[no-CD MAC] (e.g. \cite{efficient_emulation_of_single_hop_network, CZUMAJ20192})
	In this model, each station $v$ can detect two communication states: \textit{Transmission} when exactly one station from $N^+(v)$ transmitted and \textit{Noise} in any other case - including also the case when no station transmits - we cannot discern interfering signals from ambient noise.

	\item[CD MAC] (e.g. \cite{CZUMAJ20192})
	This model allows station $v$ to detect one of three states: \textit{Transmission}, when precisely one station from $N^+(v)$ transmits, \textit{Silence} if no station transmits and \textit{Noise} in other cases. It is the model closer to modern wireless communication solutions, which allows differentiating between these states.
	
	\item[Direct Messaging] (e.g. \cite{distributed_computing_a_locality_sensitive_approach})
	
	In this model, each station $v$ can send direct messages to any station $u$ if link $\{u, v\}$ exists in $E$, and each such station $u$ can detect the message coming distinctively from node $v$. In particular, a station in a given round can receive a message from all its neighbors. No collisions occur in this model. 
	
\end{description}

In the Beeping Model, in principle, we assume the signal represents a single bit (present or absent signal), while in the other models, one can assume that the messages are more complex and contain many bits. That is, during a single slot with "Transmission," many bits can be transmitted \footnote{Typically, it is assumed that the communication channel allows in a single slot to transmit at least a unique identifier of a station with $\Theta(\log n)$ bits, where $n$ is the number of stations.}. 

Note that other, less popular models can also be possible and naturally motivated by some real-life networks (e.g., systems where the collision occurs starting from some threshold of the number of transmitting stations. Below this threshold, the channel capacity allows the correct delivery of all messages.

The example of several transmission rounds is presented in \autoref{fig:execution_example}. There are six rounds, and in each, different stations are transmitting. In \autoref{tab:execution_example_data}, the observable channel states from the station $v_0$ are presented for each of the described model types.

\begin{figure}[h]
	\centering
	\begin{subfigure}[t]{0.31\textwidth} \centering
		\begin{tikzpicture}[]
		\coordinate [label=left:\small$v_0$] (S0) at (0, 0);
		\coordinate [label=right:\small$v_1$] (S1) at (0.75, 1);	
		\coordinate [label=below:\small$v_2$] (S2) at (0.75, -1);
		\coordinate [label=right:\small$v_3$] (S3) at (1.5, 0);
		
		\fill [blue] (S0) circle (0.9mm);
		\fill [white] (S0) circle (0.8mm);
		
		\fill [red] (S2) circle (0.8mm);
		\fill [red] (S3) circle (0.8mm);
		
		\fill [black] (S0) circle (0.5mm);
		\fill [black] (S1) circle (0.5mm);
		\fill [black] (S2) circle (0.5mm);
		\fill [black] (S3) circle (0.5mm);
		
		\draw [shorten >= 0.2cm, shorten <=0.2cm] (S0) -- (S1);
		\draw [shorten >= 0.2cm, shorten <=0.2cm] (S0) -- (S2);
		\draw [shorten >= 0.2cm, shorten <=0.2cm] (S0) -- (S3);
		\draw [shorten >= 0.2cm, shorten <=0.2cm] (S2) -- (S3);
		
		\draw [red, opacity=0.5, line width=0.5mm, shorten >= 0.18cm, shorten <=0.18cm] (S0) -- (S2);
		\draw [red, opacity=0.5, line width=0.5mm, shorten >= 0.18cm, shorten <=0.18cm] (S0) -- (S3);
		\draw [red, opacity=0.5, line width=0.5mm, shorten >= 0.18cm, shorten <=0.18cm] (S2) -- (S3);
		\end{tikzpicture}
		\caption{Round I.}
	\end{subfigure}
	\hfill
	\begin{subfigure}[t]{0.31\textwidth} \centering
		\begin{tikzpicture}[]
		\coordinate [label=left:\small$v_0$] (S0) at (0, 0);
		\coordinate [label=right:\small$v_1$] (S1) at (0.75, 1);	
		\coordinate [label=below:\small$v_2$] (S2) at (0.75, -1);
		\coordinate [label=right:\small$v_3$] (S3) at (1.5, 0);
		
		\fill [blue] (S0) circle (0.9mm);
		\fill [white] (S0) circle (0.8mm);
		
		\fill [black] (S0) circle (0.5mm);
		\fill [black] (S1) circle (0.5mm);
		\fill [black] (S2) circle (0.5mm);
		\fill [black] (S3) circle (0.5mm);
		
		\draw [shorten >= 0.2cm, shorten <=0.2cm] (S0) -- (S1);
		\draw [shorten >= 0.2cm, shorten <=0.2cm] (S0) -- (S2);
		\draw [shorten >= 0.2cm, shorten <=0.2cm] (S0) -- (S3);
		\draw [shorten >= 0.2cm, shorten <=0.2cm] (S2) -- (S3);
		\end{tikzpicture}
		\caption{Round II.}
	\end{subfigure}
	\hfill
	\begin{subfigure}[t]{0.31\textwidth} \centering
		\begin{tikzpicture}[]
		\coordinate [label=left:\small$v_0$] (S0) at (0, 0);
		\coordinate [label=right:\small$v_1$] (S1) at (0.75, 1);	
		\coordinate [label=below:\small$v_2$] (S2) at (0.75, -1);
		\coordinate [label=right:\small$v_3$] (S3) at (1.5, 0);
		
		\fill [blue] (S0) circle (0.9mm);
		\fill [white] (S0) circle (0.8mm);
		
		\fill [red] (S2) circle (0.8mm);
		
		\fill [black] (S0) circle (0.5mm);
		\fill [black] (S1) circle (0.5mm);
		\fill [black] (S2) circle (0.5mm);
		\fill [black] (S3) circle (0.5mm);
		
		\draw [shorten >= 0.2cm, shorten <=0.2cm] (S0) -- (S1);
		\draw [shorten >= 0.2cm, shorten <=0.2cm] (S0) -- (S2);
		\draw [shorten >= 0.2cm, shorten <=0.2cm] (S0) -- (S3);
		\draw [shorten >= 0.2cm, shorten <=0.2cm] (S2) -- (S3);
		
		\draw [red, opacity=0.5, line width=0.5mm, shorten >= 0.18cm, shorten <=0.18cm] (S0) -- (S2);
		\draw [red, opacity=0.5, line width=0.5mm, shorten >= 0.18cm, shorten <=0.18cm] (S2) -- (S3);
		\end{tikzpicture}
		\caption{Round III.}
	\end{subfigure}
	\hfill
	\begin{subfigure}[t]{0.31\textwidth} \centering
		\begin{tikzpicture}[]
		\coordinate [label=left:\small$v_0$] (S0) at (0, 0);
		\coordinate [label=right:\small$v_1$] (S1) at (0.75, 1);	
		\coordinate [label=below:\small$v_2$] (S2) at (0.75, -1);
		\coordinate [label=right:\small$v_3$] (S3) at (1.5, 0);
		
		\fill [blue] (S0) circle (0.9mm);
		\fill [white] (S0) circle (0.8mm);
		
		\fill [red] (S1) circle (0.8mm);
		\fill [red] (S2) circle (0.8mm);
		
		\fill [black] (S0) circle (0.5mm);
		\fill [black] (S1) circle (0.5mm);
		\fill [black] (S2) circle (0.5mm);
		\fill [black] (S3) circle (0.5mm);
		
		\draw [shorten >= 0.2cm, shorten <=0.2cm] (S0) -- (S1);
		\draw [shorten >= 0.2cm, shorten <=0.2cm] (S0) -- (S2);
		\draw [shorten >= 0.2cm, shorten <=0.2cm] (S0) -- (S3);
		\draw [shorten >= 0.2cm, shorten <=0.2cm] (S2) -- (S3);
		
		\draw [red, opacity=0.5, line width=0.5mm, shorten >= 0.18cm, shorten <=0.18cm] (S0) -- (S1);
		\draw [red, opacity=0.5, line width=0.5mm, shorten >= 0.18cm, shorten <=0.18cm] (S0) -- (S2);
		\draw [red, opacity=0.5, line width=0.5mm, shorten >= 0.18cm, shorten <=0.18cm] (S2) -- (S3);
		\end{tikzpicture}
		\caption{Round IV.}
	\end{subfigure}
	\hfill
	\begin{subfigure}[t]{0.31\textwidth} \centering
		\begin{tikzpicture}[]
		\coordinate [label=left:\small$v_0$] (S0) at (0, 0);
		\coordinate [label=right:\small$v_1$] (S1) at (0.75, 1);	
		\coordinate [label=below:\small$v_2$] (S2) at (0.75, -1);
		\coordinate [label=right:\small$v_3$] (S3) at (1.5, 0);
		
		\fill [blue] (S0) circle (0.9mm);
		\fill [white] (S0) circle (0.8mm);
		
		\fill [red] (S0) circle (0.8mm);
		\fill [red] (S2) circle (0.8mm);
		
		\fill [black] (S0) circle (0.5mm);
		\fill [black] (S1) circle (0.5mm);
		\fill [black] (S2) circle (0.5mm);
		\fill [black] (S3) circle (0.5mm);
		
		\draw [shorten >= 0.2cm, shorten <=0.2cm] (S0) -- (S1);
		\draw [shorten >= 0.2cm, shorten <=0.2cm] (S0) -- (S2);
		\draw [shorten >= 0.2cm, shorten <=0.2cm] (S0) -- (S3);
		\draw [shorten >= 0.2cm, shorten <=0.2cm] (S2) -- (S3);
		
		\draw [red, opacity=0.5, line width=0.5mm, shorten >= 0.18cm, shorten <=0.18cm] (S0) -- (S1);
		\draw [red, opacity=0.5, line width=0.5mm, shorten >= 0.18cm, shorten <=0.18cm] (S0) -- (S2);
		\draw [red, opacity=0.5, line width=0.5mm, shorten >= 0.18cm, shorten <=0.18cm] (S0) -- (S3);
		\draw [red, opacity=0.5, line width=0.5mm, shorten >= 0.18cm, shorten <=0.18cm] (S2) -- (S3);
		\end{tikzpicture}
		\caption{Round V.}
	\end{subfigure}
	\hfill
	\begin{subfigure}[t]{0.31\textwidth} \centering
		\begin{tikzpicture}[]
		\coordinate [label=left:\small$v_0$] (S0) at (0, 0);
		\coordinate [label=right:\small$v_1$] (S1) at (0.75, 1);	
		\coordinate [label=below:\small$v_2$] (S2) at (0.75, -1);
		\coordinate [label=right:\small$v_3$] (S3) at (1.5, 0);
		
		\fill [blue] (S0) circle (0.9mm);
		\fill [white] (S0) circle (0.8mm);
		
		\fill [red] (S1) circle (0.8mm);
		
		\fill [black] (S0) circle (0.5mm);
		\fill [black] (S1) circle (0.5mm);
		\fill [black] (S2) circle (0.5mm);
		\fill [black] (S3) circle (0.5mm);
		
		\draw [shorten >= 0.2cm, shorten <=0.2cm] (S0) -- (S1);
		\draw [shorten >= 0.2cm, shorten <=0.2cm] (S0) -- (S2);
		\draw [shorten >= 0.2cm, shorten <=0.2cm] (S0) -- (S3);
		\draw [shorten >= 0.2cm, shorten <=0.2cm] (S2) -- (S3);
		
		\draw [red, opacity=0.5, line width=0.5mm, shorten >= 0.18cm, shorten <=0.18cm] (S0) -- (S1);
		\end{tikzpicture}
		\caption{Round VI.}
	\end{subfigure}
	\caption{\label{fig:execution_example}Example of an algorithm execution in multi-hop network. Dots marked by red denote the transmitting stations, red lines mark \textit{active} links, and the blue marked node $v_0$ is the analyzed receiver.}
\end{figure}
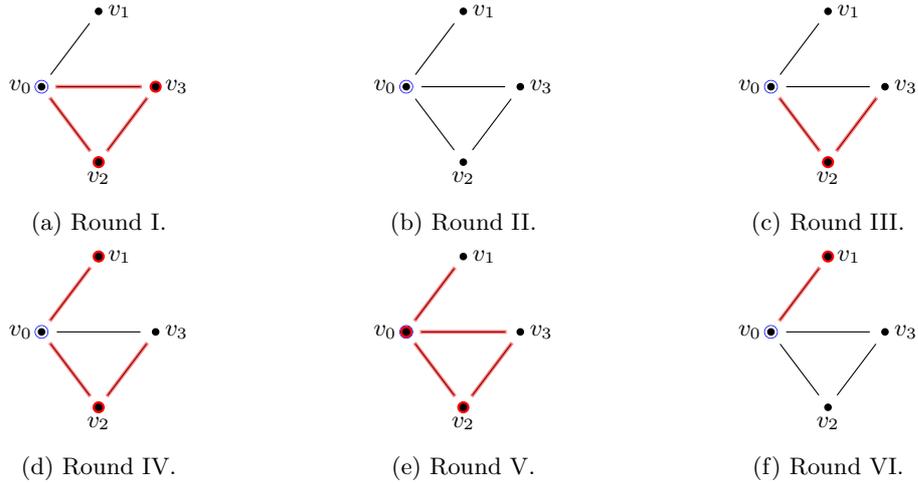

\begin{figure}[h]
	\centering \small
	\begin{tabular}{ |r|c|c|c|c|c|c|}
		\hline
		\textit{Round:} & \textit{I} & \textit{II} & \textit{III} & \textit{IV} & \textit{V} & \textit{VI} \\
		\hline
		\textbf{Beeping model} & \textit{Beep} & \textit{Silence} & \textit{Beep} & \textit{Beep} & \textit{None} & \textit{Beep} \\
		\hline
		\textbf{no-CD MAC} & \textit{Noise} & \textit{Noise} & \textit{Transmission} & \textit{Noise} & \textit{None} & \textit{Transmission} \\
		\hline
		\textbf{CD MAC} & \textit{Collision} & \textit{Silence} & \textit{Transmission} & \textit{Collision} & \textit{None} & \textit{Transmission} \\
		\hline
		\textbf{Direct Messaging} & $m_{2, 0}, m_{3,0}$ & $\emptyset$ & $m_{2, 0}$ & $m_{1, 0}, m_{2,0}$ & \textit{None} & $m_{1, 0}$ \\
		\hline
	\end{tabular}
	\caption{\label{tab:execution_example_data}Description of channel states observable by station $v_0$ from \autoref{fig:execution_example} at each round, for different channel types.}
\end{figure}

\subsection{Adversary model}~\label{q2Ref}
We can imagine it as some spying entity located close to the wireless network, capable of detecting limited information about the communication in the network. More precisely, in each round, the adversary gains some feedback from the network's communication. The feedback is the value of a feedback function for a given state (transmissions of all stations) of the network in a given slot. We will analyze a few types of adversaries modeled as \textit{feedback functions}.
\begin{enumerate}
	\item \label{adv_model_0} \textbf{Beep detecting adversary} - in each round, the adversary can detect if at least one station is transmitting.
	\item \label{adv_model_1} \textbf{Transmission counting adversary} - in each round, the adversary can detect how many stations are transmitting.
	\item  \label{adv_model_2} \textbf{Local adversary} - states of the local channels of a subset of stations are presented to the adversary.  
	\item \label{adv_label_3} \textbf{Full information adversary} - the adversary gains knowledge about all the communication (but does not know the content of the transmitted messages). 
\end{enumerate}

The information that the adversary received will be in the form of a stream $s \in \NN^*$, e.g., stream $(1, 0, 2, 0, 1)$ will mean that in the first round, only one station transmitted, in second round none station transmitted, in third - two stations transmitted and so on. Obviously, in adversary model \ref{adv_model_0}, it will be limited to $s \in \{0, 1\}^*$.

In \autoref{tab:execution_example_adversary}, we present what different adversaries can see, given the execution of the algorithm from \autoref{fig:execution_example}.

\begin{figure}[h]
	\centering \small
	\begin{tabular}{ |r|p{1.5cm}|c|c|c|c|c|}
		\hline
		\textit{Round:} & \centering \textit{I} & \textit{II} & \textit{III} & \textit{IV} & \textit{V} & \textit{VI} \\
		\hline
		\textbf{Beep detecting adversary} & \centering \textit{Beep} & \textit{Silence} & \textit{Beep} & \textit{Beep} & \textit{Beep} & \textit{Beep} \\
		\hline
		\textbf{Transmission counting adversary} & \centering \textit{2} & \textit{0} & \textit{1} & \textit{2} & \textit{2} & \textit{1} \\
		\hline
	\end{tabular}
	\caption{\label{tab:execution_example_adversary} Different data acquired from rounds of execution presented in \autoref{fig:execution_example} by different types of adversaries.}
\end{figure}

\noindent The \textbf{Full information adversary} detects the following information per each round of algorithm from \autoref{fig:execution_example}:

\renewcommand*\descriptionlabel[1]{\hspace\labelsep\normalfont #1}
\begin{description}
	\item [Round I:] $m_{2, 0}, m_{2,3},m_{3,0}, m_{3,2}$.
	\item [Round II:] $\emptyset$.
	\item [Round III:] $m_{2, 0}, m_{2,3}$.
	\item [Round IV:] $m_{1, 0}, m_{2,0},m_{2,3}$.
	\item [Round V:] $m_{0, 1}, m_{0,2},m_{0,3}, m_{2,0}, m_{2,3}$.
	\item [Round VI:] $m_{1, 0}$.
\end{description}

We plan to present the model in a possibly general form. Thus, we assume that the adversary may have some prior knowledge about the executed protocol and the network itself. It can be modeled as a probability distribution. The adversary aims to enrich its knowledge using the feedback from the execution. Note the adversary may have no exact information about the network; however, given the feedback from the execution, some scenarios turned out to be significantly more probable. Indeed, the execution can make some scenarios (e.g., about the number of stations) a~posteriori more or less probable, even without pointing to the exact one. This case can be a security threat and must be considered in the formal model.

\subsection{Algorithm's evaluation}

In analyzing the problem of information hiding, we encountered many fundamental issues with measuring the algorithm's quality and cost (understood as additional time and energy spent to obtain new properties). 
In some models, adding extra rounds of transmission to hide the accurate execution is indefensible. Thus, completing the same task is more expensive in terms of communication as well as the total time of execution. Such an approach was presented in \cite{DBLP:conf/europar/BojkoKMS23}, where the universal algorithm for hiding the size of the network was introduced and was based on each station having a probability to simulate one additional station in the single-hop model. It was presented under the regime of \textit{size-hiding} regime, which was based on the \textit{differential privacy}, presented in \cite{algorithmic_foundations_of_differential_privacy}.

\begin{definition}\label{def3}
	(hiding property)
	Let $f_{\mathcal{A}}$ be a feedback function with values in $\mathcal{Y}$ representing the knowledge of the adversary from all the slots of the algorithm's $\mathcal{A}$ execution. Let $\mathbb{A}$ be a set of possible algorithms (including their parameters), and let $\mathbb{N}$ be the set of all possible network parameters. Let $\Xi = \mathbb{A} \times \mathbb{N}$. Moreover, let $(\Xi, d)$ be a metric space. We say that $\mathcal{A}\in \mathbb{A}$ is $(d,\Xi,l,  \varepsilon, \delta)$-\textit{hiding} when for any  $S \subset  \mathcal{Y}$:
	
	\begin{equation}\label{dpequation}
	\Pr[f_{\mathcal{A}}(x)\in S] \leq \exp(\varepsilon)\Pr[f_{\mathcal{A}}(y)\in S] + \delta
	\end{equation}
	for all $x,y \in \Xi$ such that $ d(x,y)\leq l$.
\end{definition}
Note that in the assumed model, the feedback function  (possibly randomized) depends only on  $x\in \Xi$~.
This definition is a generalized version of \textit{Definition 1} from \cite{DBLP:conf/europar/BojkoKMS23}, with $f_{\mathcal{A}}$ being the beeping function ($1$ if at least one station is transmitting, $0$ otherwise), $\mathbb{N}$ being the set of all fully connected networks that can be identified with natural numbers. Moreover, the metric is $d=|n-m|$ for all $n,m \in \mathcal{N}$.

Except for the security (hiding) property, some other metrics need to be considered while evaluating the hiding method. Similarly to the bulk of previous papers on information hiding in distributed systems, in all suggested methods, our paper is somehow based on the redundancy of communication (adding some surplus actions to obfuscate the adversary's view). In effect, the obfuscated algorithm is somehow more expensive concerning the execution time and the energy necessary for completing the algorithm. The latter can be measured as a value proportional to the maximal number of transmissions over all stations participating in the protocol. This approach is motivated by two facts:

\begin{itemize}
	\item listening is an order of magnitudes less energy consuming than transmitting; 
	\item the system's lifetime is equal to the shortest life over all stations.
\end{itemize}
Let  $\mathcal{E}(\mathcal{A}), \mathcal{T}(\mathcal{A})$ be an energy and a time of execution of an algorithm $\mathcal{A}$. By $S_{T}$, let us define all the algorithms completing a given task $T$. Moreover, let 

\begin{eqnarray*}
	e_T&=&\inf\limits_{\mathcal{A}\in S_{T}} \mathcal{E}(\mathcal{A})~.\\
	t_T&=&\inf\limits_{\mathcal{A}\in S_{T}} \mathcal{T}(\mathcal{A})~,
\end{eqnarray*}

That is, $e_T$ and $t_T$ are optimal time and energy needed to complete a task $T$, respectively. Let $S^{*,\theta}_{T}$  be the set of algorithms for task $T$ hiding the execution with respect to some model parameters $\theta$ (including 
$\delta , \varepsilon, d, l$)~. Analogously we define 

\begin{eqnarray*}
	e^{*,\theta}_T&=&\inf\limits_{\mathcal{A}\in S^{*,\theta}_{T}} \mathcal{E}(\mathcal{A})~,\\
	t^{*,\theta}_T&=&\inf\limits_{\mathcal{A}\in S^{*,\theta}_{T}} \mathcal{T}(\mathcal{A})~.
\end{eqnarray*}

As the \textit{cost of hiding} w.r.t the time of execution is defined as  $\frac{e^{*,\theta}_T}{e_T}$. Similarly   $\frac{e^{*,\theta}_T}{e_T}$ is the \textit{cost of hiding} w.r.t energy.

    \section{Taxonomy}\label{taxonomy}

Compared to the results from the paper \cite{DBLP:conf/europar/BojkoKMS23}, where a single-hop radio network was considered, the case of a multi-hop radio network  (and similar distributed systems) 
is dramatically more complex. There are many substantially different (yet still natural)  assumptions about the topology of the network and the way the stations communicate. Even more important is the power of the adversary modeled by the feedback function that describes what the adversary may observe in the run of the protocol. 
A full description of the adversary needs to cover the a~priori knowledge of the adversary about the executed algorithm and topology. Moreover, we need to specify what the ultimate aim of the adversary is - what it wants to learn from the feedback function. In this section, we list the main categories for which the varying configurations can impact the algorithm design and evaluation.

\paragraph{Network topology}\mbox{}\\
We assume that a graph with nodes representing stations describes the network topology. That is, the connection between any pair of nodes is symmetric. The signal transmitted by station $x$ reacheches station $y$ if and only if $\{x,y\}$ is an edge in the graph. Similarly,  $x$ gets the signal if $y$ transmits. We assume that the graph is connected. 
\begin{itemize}
	\item Single-hop - the network is represented by a complete graph. 
	\item Multi-hop -  at least two nodes are not connected in the graph. That is, there are at least two non-connected stations $x$ and $y$. In particular, to deliver a message between them, one needs to use a path of relay stations. Clearly, in this case, delivering a message from $x$ to $y$ takes more than a single round. 
	
\end{itemize}

\paragraph{Communication channel}
\begin{itemize}
	\item Beeping model.
	\item MAC with Collison Detection.
	\item MAC without Collison Detection.
	\item Direct messaging.
\end{itemize}
Local communication channels act as described in \autoref{qqRef}.

\paragraph{Station's topology awareness}
\begin{itemize}
	\item Stations know the topology of the network. 
	\item Stations do not know the topology. It can be collectively recovered in the course of the algorithm. 
\end{itemize}

\paragraph{Station's algorithm awareness}
\begin{itemize}
	\item Algorithm aware - stations know only its code executed locally.
	\item Algorithm knowledge restricted  - stations know the code of all stations (in particular if it is the same for all stations). 
	The local inputs, however, remain unknown. 
\end{itemize}

\paragraph{Secret sharing}
\begin{itemize}
	\item Secret capable   - from the beginning of the execution, all the stations share a secret unknown to the adversary. In particular, they can use a secret to encrypt the communication that the adversary cannot read.  
	\item Open communication - at the beginning of the algorithm's execution, the stations do not share any secret.  
\end{itemize}

\paragraph{Adversary's topology awareness}
\begin{itemize}
	\item Topology aware  - the adversary knows the specified topology of the network.
	\item Topology knowledge restricted - the adversary has no or partial knowledge about the network's topology. In particular, the adversary may know that the network is a regular graph or contains, at most, some $N$ nodes. We also allow to represent the knowledge of an adversary as a probability distribution over a set of graphs. 
\end{itemize}

\paragraph{Adversary's algorithm awareness}
\begin{itemize}
	\item Algorithm aware  - the adversary knows the exact algorithm executed by all stations; however, it does not know the inputs of the stations. 
	\item The adversary has limited knowledge of the executed algorithm. 
	In particular, the knowledge can be a distribution over a set of potential algorithms. 
\end{itemize}

\paragraph{Adversary's feedback function}\mbox{}\\
Different types of feedback functions are described in \autoref{q2Ref}. We consider:
\begin{itemize}
	\item beep detecting adversary,
	\item transmissions counting adversary,
	\item local adversary,
	\item full information adversary.
\end{itemize}

\vspace{0.5cm}
\noindent
The preliminary research suggests that choosing the factors mentioned above leads to significantly different adversary capabilities. We also observed that, consequently, for each model, one needs to apply different defense strategies. One may consider some other factors influencing both the adversary's capabilities as well as possible countermeasures. We decided, however, to restrict our attention to the most important ones in order to keep the taxonomy practical.

	\section{Algorithms}\label{algorithms}

This section presents a few elementary algorithms offering information-hiding properties for chosen models from the introduced taxonomy.  

\subsection{Naive Oblivious}
This algorithm can be applied for a relatively weak \textbf{beeping model} of the feedback adversary and the strongest \textbf{direct messaging} as a communication model. Other parameters can be fixed arbitrarily. In particular, the algorithm does not assume any shared secret (\textbf{open communication} model). Moreover, the stations do not have to know topology and can have only local knowledge about the execution. 

\paragraph{Description}

The Naive Oblivious algorithm $\mathcal{N} (\mathcal{A})$  is built on the top of any algorithm $\mathcal{A}$. We assume that messages sent by stations during the protocol are of equal size $l$, and stations know the upper bound on the execution length $N$.  Naive Oblivious $\mathcal{N} (\mathcal{A})$ works as follows:

\begin{itemize}
	\item If in the original protocol $\mathcal{A}$, in a round $1 \leq t \leq N$, station $s_i$ sends a message $m_{s_i,s_j}^n$ to $s_j$, in the modified protocol in the round $t$ in the protocol $\mathcal{N} (\mathcal{A})$ the station $s_i$ sends to $s_j$ a message $1\|m_{s_i,s_j}^t$. That is, the same message is sent, however, with a prefix $'1'$. 
	
	\item If in the round $t$ the message is \textbf{not} sent in $\mathcal{A}$, in the $\mathcal{N} (\mathcal{A})$ the station $s_i$ sends to $s_j$ the \textit{dummy} message of the length $l+1$ with zeros, only.  
	
\end{itemize}

\noindent The original messages from $\mathcal{A}$ can be easily distinguished from dummies.

\paragraph{Analysis}

The analysis of security properties is obvious. One can see that the adversary can observe only a sequence of $N$ beeps. That is, the protocol is totally oblivious. In effect, one gets $\varepsilon=\delta=0$ as the security basic parameters for any properly defined $d$ and $\Xi$. On the other hand, the stations taking advantage of the significantly more informative communication model can execute the $\mathcal{A}$.

Note that the assumption about the equal length of messages sent in the protocol can be easily bypassed using, e.g., standard padding.

\subsection{Binomial Boxes Algorithm}
The simplicity of the Naive, Oblivious algorithm was based on the fact that the adversary, having just beeping feedback, was much weaker compared to the regular stations in the network that could communicate simultaneously with all their neighbors. This section introduces the Binomial Boxes Algorithm (or BBA, for short) that can be applied to an adversary still having beeping feedback with constrained regular stations (beeping model or CD/no-CD MAC). The price of reducing the difference in capabilities of the adversary and the regular stations is the requirement that the stations need to share a common secret unknown to the adversary. Moreover, the execution of the algorithm is significantly larger in terms of time and energy and depends on the parameter determining the security level.

\paragraph{Descripion}

Each time slot of execution of a regular protocol $\mathcal{A}$ is represented by a \text{box} that consists of $n+1$ consecutive regular slots. In each box, a single \textit{true slot} is chosen uniformly in a pseudo-random random manner. Other $n$ slots are independently chosen as \textit{beep dummy} or \textit{silent dummy} with probability $1/2$. The position of the true slot and decisions if the remaining slots are silent or beep dummies are to be determined by the shared secret\footnote{This can be done straightforwardly using a chain of one-way hash functions with the secret as a seed~.}. Thus, the position of the true slot in a box and the kind of dummies are known for the stations sharing the secret but remain random for the adversary.

The execution of the protocol $\mathcal{BBA}(\mathcal{A})$ is as follows:
\begin{itemize}
	\item In the true slot of the $t$-th box of $\mathcal{BBA}(\mathcal{A})$ all the stations execute the actions of the $t$-th slot of $\mathcal{A}$; 
	\item In all beep dummy slots, all the stations transmit. 
	\item In all silent dummy slots, all the stations remain silent. 
\end{itemize}

\begin{figure}[h]
	\centering \small
	\begin{tabular}{ |r||c|c|c||c|c|c||c|c|c|||r|}
		\hline
		\textit{Box} & \multicolumn{3}{c||}{Box I} & \multicolumn{3}{c||}{Box II} & \multicolumn{3}{c|||}{Box III} &  \\
		\hline
		\textit{Slot} & 0 & 1 & \textbf{2} & \textbf{0} & 1 & 2 & 0 & \textbf{1} & 2 & Message \\
		\hline
		\hline
		\textit{Station A} & S & S & \textbf{S} & \textbf{S} & B & S & S & \textbf{S} & B & SSS \\
		\hline
		\textit{Station B} & S & S & \textbf{S} & \textbf{B} & B & S & S & \textbf{B} & B & SBB \\
		\hline
		\textit{Station C} & S & S & \textbf{B} & \textbf{S} & B & S & S & \textbf{S} & B & BSS \\
		\hline
		\textit{Station D} & S & S & \textbf{B} & \textbf{B} & B & S & S & \textbf{S} & B & BBS \\
		\hline
	\end{tabular}
	\caption{\label{tab:bba_example} Example execution of $\mathcal{BBA}$ algorithm with three boxes and three slot each. The true slot is marked in bold. Notice how stations in other slots are using the same behavior.}
\end{figure}

\paragraph{Analysis}
The correctness $\mathcal{BBA}(\mathcal{A})$ of the protocol is obvious. Since the stations neglect dummy slots, the execution of the true slots needs to give the same result\footnote{We do not formalize explicitly the results of the protocol, but it can be seen, for example, as the states of local memories of all stations.} as  $\mathcal{A}$. 

More subtler analysis of information-hiding properties is needed. 
Let us observe that the adversary can only distinguish the slot with and without any transmitting station. 

Let us call the box representing the slot with the transmission in the true slot a \textit{beep box} and the remaining a \textit{silent box}.
Since the position of the true slot in each box is random from the perspective of the adversary, the only information the adversary can learn is the number of beep slots (including the true slot) in a given box. One can easily see that statistically, there is one more beep in the beep slot. Intuitively, the difference between the types of boxes vanishes with a growing parameter $n$. 
Formally, the number of beeping slots in the silent box is binomially distributed $T_{S} \sim \mbox{Bionmial}(n,1/2)$, while in the case of beep box, we got  $T_{B} \sim  \mbox{Bionmial}(n,1/2) 
+ 1$~. 
Let us recall the following version of the Chernoff bound.  

\begin{fact}\label{nieu}
	Let $X$ be binomially distributed with parameters $n$ and $p$. 
	For any $\delta >0$ and $\mu = np$ following holds: 
	
	$$\Pr[|X-\mu|\geq \delta \mu ] \leq 2\exp\left( - \frac{\delta^2 \mu}{3} \right)~.$$
\end{fact}

This version of the Chernoff inequality is obtained by a simple union bound to unify cases with upper and lower bounds for binomial distribution (see, e.g., \cite{mulzer2019proofs}). 
Let $\delta = 2 \xi \frac{1}{\sqrt{n}}$ for some $\xi > 1$ being a security parameter. Applying directly $\delta$ to~\ref{nieu} to $T_{S}$ we get:

$$\Pr\left [|T_{S}- \frac{n}{2}|> \xi\sqrt{n} \right ]\leq 2 \exp \left ( -\frac{4}{6} \xi^2 \right) < \exp \left ( - \frac{\xi^2}{2}\right )~. $$
\noindent
It directly implies that:

$$\Pr\left [|T_{B} (-1) - \frac{n}{2}|> \xi\sqrt{n} \right ]<   \exp \left ( - \frac{\xi^2}{2}\right )~. $$

\noindent
In effect values of $T_{B}$ and $T_{S}$ are in the interval 
$\mathbf{I}=\left [ \frac{n}{2} - \xi\sqrt{n}  , \frac{n}{2} + \xi\sqrt{n} +1\right ]$ with probability exceeding $1-\exp \left ( - \frac{\xi^2}{2}\right )$~. For extreme values, it can be easy to distinguish if the result is from $T_{B}$  or $T_{S}$. For example, having beeps in all $n+1$ slots, it is evident that we deal with beep-box. We show, however, that all the values from 
$\mathbf{I}$ can appear in $T_{B}$  or $T_{S}$ almost with the same probabilities. Note that for any $ 2 \leq l \leq n$:

$$\frac{\Pr(T_{S}=l)}{\Pr(T_{B}=l)}=\frac{\Pr(T_{S}=l)}{\Pr(T_{S}=l-1)}=\frac{{n \choose l} \frac{1}{2^n} }{ {n \choose l-1} \frac{1}{2^n} }= \frac{n-l+1}{l}~:=f(n,l)~.$$

\noindent One can see that for  $l \in \mathbf{I}$ we have:
$$f(n,l)\geq \frac{n - \left ( \frac{n}{2} + \xi\sqrt{n} +1 \right) +1 }{\frac{n}{2} + \xi\sqrt{n} +1 } = 1 - \frac{2\xi\sqrt{n} +1}{\frac{n}{2} + \xi\sqrt{n} +1} \geq 1 - \frac{2\xi\sqrt{n}+1}{\frac{n}{2}}\geq 1 - \frac{5\xi}{\sqrt{n}}~. $$

\noindent In the same way, one can show that for  $l \in \mathbf{I}$ we have:
$$f(n,l)\leq 1+ \frac{7\xi}{\sqrt{n}}~.$$

Thus the ratio  $|f(n,l)|\leq 1 + x $ for some  $|x| = \Theta\left( \frac{\xi}{\sqrt{n}} \right)$ for all $l\in \mathbf{I}$. Since $\ln x = 1 + x + \Delta$ for some $|\Delta| < x^2$ if  $x<1$, we easily get that:
$$f(n,l) \leq \exp(\varepsilon)$$
for:
$$\varepsilon = \ln (|f(n,l)|) = \Theta\left( \frac{\xi}{\sqrt{n}} \right)~.$$
Finally we need to recall that we proved that $l\in \mathbf{I}$ with probability at least $1  - \exp\left(\frac{\xi^2}{2} \right)$. 
As a consequence of the above considerations, one gets the following fact. 

\begin{fact}\label{fifi2}
	Let $\mathbb{A}^{(k)}$ be a set of algorithms lasting exactly $k$ rounds in the MAC communication channel\footnote{with or without CD}.
	For any $\mathcal{A}\in \mathbb{A}^{(k)}$ the algorithm   $\mathcal{BBA}(\mathcal{A})$ with parameters $n>0$ and $0<\xi<1$ is  $(\Xi,d,l,\varepsilon,\delta)$-hiding for 
	\begin{itemize}
		\item $\varepsilon = \Theta\left( k\cdot\frac{\xi}{\sqrt{n}} \right),$
		\item $\delta =  \Theta\left(k\cdot \exp\left(-\frac{\xi^2}{2} \right)\right), $
		\item $\Xi=\mathbb{A}^{(k)}\times \mathbf{N}$ for any $\mathbf{N}$,
	\end{itemize}
	any metric $d$ and any number $l>0$~.
\end{fact}

The parameters for $k=1$ (a single-round algorithm) follow directly from the analysis described above. The case for $k>1$ is a direct consequence of the composition theorem (see eg.\cite{algorithmic_foundations_of_differential_privacy}). 
Note that using $\xi = \ln n^{\alpha}$ for $\alpha>1$  gives a reasonable trade-off between security parameters with $\delta,\varepsilon \xrightarrow[{}]{n} 0$, assuming that $k$ is fixed.
Let us also stress that $\mathcal{BBA}$ is a very general algorithm. For that reason, in the Fact~\ref{fifi2}, we can use any metric $d$ and a very general class of cases $\Xi$. In fact, this means that the $\mathcal{BBA}$ algorithm hides all the details of the algorithm and the network but the length of the execution.
Note that the above theorem can be optimized, and better results can be obtained (especially for limited types of $\mathcal{A}$ algorithm).

    \section{Related and Future Work}\label{related}

The single-hop radio networks and the problem of hiding the exact size of the network were analyzed in \cite{DBLP:conf/europar/BojkoKMS23}. Note that the multi-hop model is dramatically more complex. There are several papers concerning the multi-hop network for different types of communication channels, e.g. for beeping model \cite{deploying_wireless_networks_with_beeps, near_optimal_leader_election_in_multi_hop_radio_networks, optimal_multi_broadcast_with_beeps_using_group_testing, CZUMAJ20192}, different variants of MAC \cite{efficient_emulation_of_single_hop_network, CZUMAJ20192} and direct messaging \cite{distributed_computing_a_locality_sensitive_approach}.

The \textit{differential privacy}, which inspired this paper, was described in \cite{calibrating_noise_to_sensitivity_in_private_data_analysis, algorithmic_foundations_of_differential_privacy}. Its application for the learning algorithms boosting method was analyzed in \cite{lit:boosting_and_differential_privacy}. It was also considered for the protecting privacy of distributed systems scenarios in \cite{lit:distributed_differential_privacy} and \textit{Internet of Things} in \cite{lit:privacy_preserved_data_sharing, lit:privacy_preservation_in_blockchain}.

Despite significant effort and much research devoted to security in distributed systems, to the best of our knowledge, this type of information-hiding property has not been investigated systematically. 
We believe that the presented model with the constructed taxonomy can 
be a good foundation for analyzing the information hidden in distributed systems. The most problematic challenge is to consider that even a minor change to the model significantly affects the adversary's capabilities and reasonable defense strategies.

\end{document}